\documentclass[preprint,aps]{revtex4}

\usepackage{epsfig}
\usepackage{graphicx}
\usepackage{amssymb}
\usepackage{amsmath}

\begin{document}

\title{Weak Mixing Angle and
Higgs Mass in Gauge-Higgs Unification Models with Brane Kinetic Terms}

\author{ Jubin Park }
\email{honolo@phys.nthu.edu.tw}
\vskip 0.5cm

\affiliation{Department of Physics, National Tsing Hua University, HsinChu 300, Taiwan}

\author{Sin Kyu Kang}
\email{skkang@snut.ac.kr}
\vskip 0.5cm

\affiliation{School of Liberal Arts, Seoul-Tech., Seoul 139-743, Korea}

%\date{\today}
\vspace{2cm}
%\maketitle

\begin{abstract}
We show that the idea of Gauge-Higgs unification(GHU) can be rescued from the constraint of weak mixing angle by introducing localized brane kinetic terms in higher dimensional GHU models with bulk and simple gauge groups.
We find that those terms lead to a ratio between Higgs and W boson masses, which is a little bit deviated from the one
derived in the standard model. From numerical analysis, we find that the current lower bound on the Higgs mass tends to prefer to exceptional groups $E_{6,~7,~8}$ rather than other groups like $SU(3l)$, $SO(2n+1)$, $G_{2}$, and $F_{4}$ in 6-dimensional (D) GHU models irrespective of the compactification scales. For the compactification scale below 1 TeV, the Higgs masses in
6D GHU models with $SU(3l)$, $SO(2n+1)$, $G_{2}$, and $F_{4}$ groups are predicted to be less than the current lower bound unless a model parameter responsible for re-scaling $SU(2)$ gauge coupling is taken to be unnaturally large enough.
To see how the situation is changed in more higher dimensional GHU model, we take
7D $S^{3}/ \mathbb{Z}_{2}$ and 8D $T^{4}/Z_{2}$ models.
It turns out from our numerical analysis that these higher dimensional GHU models with gauge groups except for $E_{6}$ can lead
to the Higgs boson whose masses are predicted to be above the current lower bound  only for the compatification scale above 1 TeV  without taking unnaturally large value of the model parameter, whereas the Higgs masses in the GHU models with $E_{6}$
are compatible with the current lower bound even for the compatification scale below 1 TeV.  \\

% \pacs{PACS numbers: 11.10.Kk, 11.15.-q, 12.10.-g }
% 11.10.Kk Field theories in dimensions other than four
% 11.15.-q Gauge field theories
% 12.10.-g Unified field theories and models

\end{abstract}

\maketitle

%\end{titlepage}

%\voffset 0cm

%\begin{multicols}{2}
%\narrowtext

%\tightenlines

%%%%%%%%%%%%%%%%%%%%%%%%%%%%%%%%%%%%%%%%%%%%%%%%%%%%%%%%%%%%%%%%%%%%%%%%
\section{Introduction}
\label{sec:1}
%%%%%%%%%%%%%%%%%%%%%%%%%%%%%%%%%%%%%%%%%%%%%%%%%%%%%%%%%%%%%%%%%%%%%%%%

Much attention has been drawn to the hierarchy problem associated with Higgs boson at the quantum level.
Many interesting models including supersymmetric standard model~\cite{Dimopoulos:1981yj} and extra dimensional models~\cite{ArkaniHamed:1998rs} have been proposed on the way to solve it. The Higgs field, the only undiscovered piece of the SM, whose discovery is anticipated to be just round the corner would provide us with an essential guide to probe possible solutions of the hierarcy problem as well as the origin of mass generation.
%Unless we could find the fundamental (or final) theory so as to understand the origin of huge hierarchy between the planck scale and electroweak scale (
%or at least little hierarchy between electroweak scale and  a few TeV), the Higgs particle in the Standard Model(SM) will be remained as one uncomfortable %piece as ever in theoretical and experimental consideration. Especially great attention has been shown to the question of the hierarchy problem of the Higgs at %the quantum level. Many interesting models like minimal supersymmetric model are introduced as popular solutions of it.
%In more detail at the quantum loop level the notorious quadratic divergence of the Higgs destroys the stability of the electroweak scale if we do not fine-tune %parameters unnaturally. The UV cutoff dependency of a given theory to the Higgs mass is well-known as the "(gauge or little) hierarchy problem". Moreover the %mechanism for generating masses of matter fields and the secret of the correct electroweak symmetry breaking(EWSB) that is closely related to the Higgs, also %still remain mysteriously.
From a new philosophical point of view, the Gauge-Higgs unification(GHU) models~\cite{Manton:1979kb} have nowadays drawn renewed attention among extra dimensional models as one interesting alternative solution of the hierarchy problem and the origin of the electroweak symmetry breaking(EWSB). In the GHU scheme, the Higgs field can be identified as an internal component of a higher dimensional gauge field of a group involving electroweak gauge group, and thus the disastrous quadratic divergences associated with the Higgs field are not generated due to the higher dimensional gauge symmetry. In addition, the EWSB can nicely be accomplished by suitable boundary conditions on orbifold and/or quantum fluctuation using Wilson loop on non-simply connected extra dimensional space called Hosotani mechanism~\cite{Hosotani:1988bm}.

%In fact the original idea of the GHU traces back to a pioneer work of Manton in 1979~\cite{Manton:1979kb}. He considered 6-dimensional(6D) pure gauge theory in usual Minkowski spacetime with an extra 2-dimensional(2D) sphere $S^{2}$ in order to embed the Weinberg-Salam model at the electroweak scale into it naturally. Because the extra compactified dimension could not reduce the rank of given gauge group at that time, in the paper he discussed the only 3 possible choices ($SU(3)$, $O(5)$ and $G_{2}$) and showed that a 4-dimensional(4D) effective lagrangian with some scalar particles and its potential, can be naturally derived from the 6D pure gauge theory with one special monopole background in extra dimension. He also found one interesting relation that the mass of Higgs boson can be estimated from a mass of W boson and the mixing angle given by the model. Really the natural appearance of scalar particle and its potential from the higher dimensional gauge theory could make this GHU idea more attractive. Since this work, these common features of GHU models have been main ingredients for building a more realistic model so far.

In spite of such nice features, as shown by Aranda and Wudka \cite{Grzadkowski:2006tp,Aranda:2010hn}, any GHU models with bulk (and simple) gauge symmetry can not predict correct weak mixing angle $\theta_W (\sim \pi/6)$, so called the Weinberg angle, measured from experiments.
Except for the weak mixing angle, the GHU models are consistent with other low-energy data.
%Since the GHU models are constructed based on one large gauge group involving SM weak gauge group $SU(2)_{L} \times U(1)_{Y}$, \textbf{the tree level value of the weak mixing angle is perfectly determined by group theoretical analysis depending on the position of the generator that can play a role of the Higgs particle in root space. However in general these values are deviated from the measured value from experiments.}
To remedy this serious problem in the GHU models, three categories of the solutions have been proposed;
$1.$ adding the localized brane kinetic terms~\cite{Burdman:2002se}, $2.$ incorporating (anomalous) additional $U(1)$ gauge symmetry~\cite{Antoniadis:2001cv}, $3.$ imposing violation of higher dimensional Lorentz symmetry $SO(1,4)$~\cite{Panico:2005dh}.

In this paper, we consider the first possibility to resolve the problem for the weak mixing angle, and particularly show how the correct value of the weak mixing angle can be obtained without substantial modification of the models by re-scaling $SU(2)$ and $U(1)$ couplings, respectively.
We obtain a little bit modified ratio between Higgs scalar and W boson masses, compared
 with that predicted in the SM, and the Higgs mass is shown to be predicted as a function of a model parameter which is responsible for
 re-scaling of $SU(2)$ gauge coupling for the values of the weak mixing angle and W boson mass taken to be their experimental values.
%It is interesting that these numerical results can explain why we need at least these two free parameters(we call these parameters $c_{1}$ and $c_{2}$) for re-scaling from the well-known Higgs mass exclusion bound.}
The paper is organized as follows. In section II, we briefly introduce a toy $SU(3)$ model in order to see how the brane kinetic terms in GHU models affect the gauge couplings of $SU(2)$ and $U(1)$. In section III, we present several experimental constraints with which the GHU models should be satisfied and we show which GHU models can be phenomenologically viable and investigate the effects of brane kinetic terms in those
models. We also estimate the Higgs mass in several specific extra dimensions for given compactification scales, and derive the ratio between the Higgs scalar and $W$ boson masses in terms of the mixing angle and two model parameters responsible for
re-scaling $SU(2)$ and $U(1)$ gauge couplings.
In section IV, we discuss our numerical results and conclude this paper with some implication.

\section{5-dimensional SU(3) Gauge-Higgs unification model with brane kinetic terms as a toy model}
\label{sec:2}
%%%%%%%%%%%%%%%%%%%%%%%%%%%%%%%%%%%%%%%%%%%%%%%%%%%%%%%%%%%%%%%%%%%%%%%%
In this section, we investigate the effects of brane kinetic terms in 5-dimensional(D) gauge-Higgs unification model with SU(3) gauge symmetry as a toy model~\cite{Bhattacharyya:2009gw}.
Let us begin by considering a lagrangian with a higher dimensional SU(3) gauge symmetry on a $S_{1}/ \mathbb{Z}_{2}$ orbifold in 5 dimensions without brane kinetic terms,
\begin{equation}
\mathcal{L}_{5D}=\int{d^{4}x} \int{dy} -\frac{1}{4} \big(F^{a}_{MN}\big)^{2}~,
\end{equation}
where $F_{MN}^{a}$ is the field strength denoted by $=\partial_{M}A_{N}^{a}-\partial_{N}A_{M}^{a}+g_{5D}f^{abc}A_{M}^{b}A_{N}^{c}$.
 Here, Latin indices are used to denote 5-dimensional(D) space-time coordinate, for instance $M=$ 0,1,2,3,5, whereas Greek indices represent usual 4D space-time, $\mu=$ 0,1,2,3, and here upper index $a$  indicates a group index.
The gauge symmetry $G=SU(3)$ is broken down to subgroup $H=SU(2)_{W} \times U(1)_{Y}$ due to orbifold boundary conditions imposed to the gauge fields at around $1/R$ compactification scale with a projection matrix $P=\mathrm{diag}(-1,-1,+1)$,
\begin{equation}
A_{\mu}(x,y)=P^{-1}A_{\mu}(x,-y)P,~~~~~~ A_{5}(x,y)=-P^{-1}A_{5}(x,-y)P~~~,
\end{equation}
where all $A_{\mu}$ and $A_{5}$ fields are Lie-algebra valued functions of $A_{M}=A_{M}^{a}\frac{\lambda ^{a}}{2}$, and $\lambda ^{a}$ are Gell-Mann matrices.

The components of the $A_{\mu}$ corresponding to zero modes with their generators are given by,
\begin{equation}
A_{\mu}^{(0)}=\frac{1}{2}
\left(
  \begin{array}{ccc}
    A^{3}_{\mu}+\frac{1}{\sqrt{3}}A^{8}_{\mu} & A^{1}-iA^{2}_{\mu} & 0 \\
    A^{1}+iA^{2}_{\mu} & -A^{3}_{\mu}+\frac{1}{\sqrt{3}}A^{8}_{\mu} & 0 \\
    0 & 0 & -\frac{2}{\sqrt{3}}A^{8}_{\mu} \\
  \end{array}
\right)~,
\end{equation}
where the superscript $(0)$ of $A_{\mu}$ means zero modes.
Similarly, the $A_{5}$ is given by the following form,
\begin{equation}
A_{5}^{(0)}=\frac{1}{2}
\left(
  \begin{array}{ccc}
    0 & 0 & A^{4}_{5}+iA^{5}_{5} \\
    0 & 0 & A^{6}_{5}+iA^{7}_{5} \\
    A^{4}_{5}-iA^{5}_{5} & A^{6}_{5}-iA^{7}_{5} & 0 \\
  \end{array}
\right)~.
\end{equation}
Note that each component of the $A_{5}$ has opposite $Z_{2}$ parity to the $A_{\mu}$.
Adopting familiar SU(2) notation of subgroup $H$, all components of the zero modes can be rewritten by
\begin{equation}
A_{\mu}^{(0)}+A_{5}^{(0)}=\frac{1}{2}
\left(
  \begin{array}{ccc}
    W^{3}_{\mu}+\frac{1}{\sqrt{3}}B^{8}_{\mu} & \sqrt{2}\,W^{+}_{\mu} & \sqrt{2}\,H^{*}_{5} \\
    \sqrt{2}\,W_{\mu}^{-} & -W^{3}_{\mu}+\frac{1}{\sqrt{3}}B^{8}_{\mu}  & \sqrt{2}\,H^{0}_{5} \\
    \sqrt{2}\,H^{-}_{5} & \sqrt{2}\,H^{*0}_{5} & -\frac{2}{\sqrt{3}}B^{8}_{\mu} \\
  \end{array}
\right)
\end{equation}
Let us focus on the 5D Lagrangian corresponding to the zero modes,
\begin{equation}
\mathcal{L}_{5D}=\int d^{4}x \int^{\pi R}_{0} dy -\frac{1}{4}\big( F^{a(0)}_{\mu\nu} \big)^{2}+\cdots,
\end{equation}
where we have used the fact that the fundamental domain of orbifold is $[0,\pi R]$, and many other terms including
$\partial_{\mu} A^{a(0)}_{5}$ can be ignored by taking thin brane approximation and thanks to the properties of fields under orbifold projection~\cite{Carena:2002me}.
After integrating out 5D space, the 5D lagrangian becomes
\begin{equation}
\int d^{4}x -\frac{1}{4}\big( F^{a(0)}_{\mu\nu} \big)^{2}\cdot \mathrm{Z}_{0}^2~,
\end{equation}
where $\mathrm{Z}_{0}^2\equiv{\pi R}$, and the factor contains the length (or area or volume in more higher dimension) of the fundamental domain.
Re-scaling the zero mode of the 4D gauge field, $A^{(0)a}_{\mu} \rightarrow \mathrm{Z}_{0} A^{(0)a}_{\mu}$,
and comparing the 4D effective lagrangian with normal 4D lagrangian for gauge subgroup H,
we can obtain the relation between 4D and 5D gauge coupling constants,
\begin{equation}
g_{4D}=\frac{g_{5D}}{\mathrm{Z_{0}}}=\frac{g_{5D}}{\sqrt{\pi R}}~.
\end{equation}

Now we introduce the brane kinetic terms in the 5D lagrangian,
\begin{equation}
\mathcal{L}_{B.K}= \int d^{4}x \int dy~ -\frac{1}{4}\,\delta(y)\Big[ c_{1}(F_{\mu\nu}^{a})^{2} + c_{2}(F_{\mu\nu}^{b})^{2} \Big]~,
\end{equation}
where $a$ and $b$ represent the generators of $SU(2)$ and $U(1)$ gauge group respectively, that is, $a=1,2,3$ and $b=8$.
Then, the new effective 4D Lagrangian of $SU(2)$ gauge part is given by
\begin{equation}
\mathcal{L}_{eff.}=(\mathrm{Z}_{0}^{2} + c_{1})(-\frac{1}{4}\big( F^{a(0)}_{\mu\nu} \big)^{2})~.
\end{equation}
After re-scaling followed by the same way shown above, we obtain a little bit different relation compared to the previous one,
\begin{equation}
g_{4D,~SU(2)}^{\prime}=\frac{g_{5D}}{\sqrt{\mathrm{Z}_{0}^{2}+c_{1}}}=\frac{g_{5D}}{\mathrm{Z}_{0}}
\frac{1}{\sqrt{1+\frac{c_{1}}{\mathrm{Z}_{0}^{2}}}}=\frac{g_{4D}}{\sqrt{\mathrm{Z}_{1}}}~,
\end{equation}
where we introduced a new constant factor $Z_{1}$ like usual renormalization constant in the last equality. Similarly, we can also derive a relation between two gauge couplings for $U(1)$ part as follows,
\begin{equation}
g_{4D,~U(1)}^{\prime}=\sqrt{3}\frac{g_{4}}{\sqrt{\mathrm{Z}_{2}}}~,
\end{equation}
and each $\mathrm{Z}_{i}$ is given by
\begin{equation}
\mathrm{Z}_{i}=1+\frac{c_{i}}{\mathrm{Z}_{0}^{2}}~,~~\mathrm{(i=1~or~2)}~.
\end{equation}
It is worthwhile to notice that those brane kinetic terms can modify each gauge coupling of $SU(2)$ and $U(1)$
 by the re-scaling of the gauge fields. In fact, the numerical factor $\sqrt{3}$ comes from the structure constant of $SU(3)$, and in general we can calculate this important number by using group theoretic analysis depending on symmetry breaking patterns of given models. Finally the 4D effective Lagrangian is written by
\begin{eqnarray}
\mathcal{L}_{4D}&=&-\frac{1}{4}\big( F^{a(0)}_{\mu\nu} \big)^{2} + -\frac{1}{4}\big( F^{a(8)}_{\mu\nu} \big)^{2} \nonumber \\
&+& \Big| \Big( \partial_{\mu} -i \frac{g_{4}}{\sqrt{\mathrm{Z}_{1}}} W^{a}_{\mu}T^{a} -i \frac{g_{4}}{\sqrt{\mathrm{Z}_{2}}}\sqrt{3}\,B_{\mu}Y \Big) H \Big|^{2} ~.
\end{eqnarray}
Therefore, we can identify $g_{4}/\sqrt{Z_{1}}$ as the gauge coupling $g$ of $SU(2)$ and $\sqrt{3}\,g_{4}/\sqrt{Z_{2}}$ as the gauge coupling $g^{\prime}$ of $U(1)$.
The weak mixing angle is then presented by
\begin{equation}
\tan \theta_{W} = \frac{g^{\prime}}{g} = \sqrt{3}\,\sqrt{\frac{Z_{1}}{Z_{2}}}~.
\end{equation}
In the case of $c_{1}=c_{2}=0$ (namely, $Z_{1}=Z_{2}=1$), the value of $\tan \theta_{W}$ becomes $\sqrt{3}$ which is one of the well-known results for the $SU(3)$ GHU model.

Before ending up this section, we shortly give a comment on two problems existed in the 5D $SU(3)$ GHU model. First, since any Higgs potential can not be generated at tree level due to the higher dimensional gauge symmetry and absence of the quartic couplings, a mechanism should be contrived to construct suitable Higgs potential. Second, even though some mechanisms work correctly to generate it, the size of the quantum corrections may not be enough to generate a few hundred GeV scale of top quark and the Higgs scalar masses.
In the next sections, we will show how those two problems can be naturally solved when we extend space-time to 6 dimensions(or more higher dimensions) in the GHU scheme.

%%%%%%%%%%%%%%%%%%%%%%%%%%%%%%%%%%%%%%%%%%%%%%%%%%%%%%%%%%%%%%%%%%%%%%%%
\section{phenomenologically viable Gauge-Higgs unification models}
\label{sec:3}
%%%%%%%%%%%%%%%%%%%%%%%%%%%%%%%%%%%%%%%%%%%%%%%%%%%%%%%%%%%%%%%%%%%%%%%%
\subsection{ Basic setup of GHU models and conventions of Lie algebra}

We consider general GHU models with arbitrary gauge group $G$ defined by 4+n space time, $M^{3,1} \times (\mathbb{R}^{n}/\Gamma)$, with usual 4D Minkowskian manifold  $M^{3,1}$ and  a discrete group $\Gamma$
 acting on extra dimension $\mathbb{R}^{n}$, whose  Lagrangian is given by \cite{Grzadkowski:2006tp,Aranda:2010hn},
\begin{equation}
\mathcal{L}=\int{d^{\,4+n}x} \Big\{-\frac{1}{4} \big(F^{a}_{MN}\big)^{2} \Big\}~.
\end{equation}
Here, we assume that the actions of 4+n dimensional gauge field $A^{M,a}$ under $\Gamma$ operations on $\mathbb{R}^{n}$ behave as follows;
\begin{eqnarray}
A^{\mu}_{a}(x,y') = \mathbb{V}(\gamma)_{ab}\, A^{\mu}_{b}(x,y),&&
~ A^{5}_{a}(x,y') = \mathbb{V}(\gamma)_{ab}\, A^{5}_{b}(x,y) \\
%~ A^{5}_{a}(x,y') = \mathbb{V}(\gamma)_{ab} \mathfrak{R}(\gamma) A^{5}_{b}(x,y) \\
&&y'=\gamma y=r y + l, ~~~ y \in \mathbb{R}^{n}
\end{eqnarray}
where $\gamma \equiv \{r|l\},~~\gamma \in \Gamma$, $r$ and $l$ represent rotation and translation vectors, respectively.
In order for the Lagrangian to be invariant under general operations including reflection, rotation, translation on extra compactified spaces, the matrices $\mathbb{V}$ must satisfy following properties:
\begin{equation}
f_{a'b'c'}=f_{abc} \mathbb{V}(\gamma)_{aa'}\mathbb{V}(\gamma)_{bb'}\mathbb{V}(\gamma)_{cc'},~~ \mathbb{V}^{\dagger}\mathbb{V}=1~.
\end{equation}
 Since there exist two eigenvalues $(+1$ or $-1)$ of $\mathbb{V}$ under orbifold projection $\{ -1|0 \}$, we can divide all operators into two types, $P$ corresponding to the eigenvalue $+1$ and $N$ corresponding to -1.

It is also known that the zero modes of $A_{\mu}$ and $A_{5}$ that should have Neumann boundary conditions under the translation, $y^{'} \sim y + l$, must satisfy~\cite{Grzadkowski:2005wp}
\begin{equation}
A_{\mu}^{a}=\mathbb{V}\big(\{0|\,l\}\big)_{ab} A_{\mu}^{b},~~A_{5}^{a}=\mathbb{V}\big(\{0|\,l\}\big)_{ab} A_{5}^{b}~.
\end{equation}
From these relations, we see that the zero modes of $A_{\mu}$ and $A_{5}$ have the eigenvalue $+1$ under the above translation. Combining this operator with the orbifold projection, we introduce two operators $P^+$ and $N^+$ with the index $+$ implying
eigenvalue $+1$ for the above translation.
Then, the commutation relations~\cite{Grzadkowski:2006tp} among generators are given by
\begin{equation}\label{eq:NP}
[N^{+},P^{+}] \subset N^{+},~[N^{+},N^{+}] \subset P^{+},~[N^{+},R] \subset R~,
\end{equation}
where $R$ is a set of remaining generators of $G$.
For a 5D $SU(3)$ toy model on an orbifold $S^{1}/\mathbb{Z}_{2}$,
\begin{eqnarray}
&&\mathbb{V}\big(\{0|\pi R\}\big)_{ab}=\delta_{ab},~~\mathbb{V}\big(\{-1|0\}\big)_{ab}=\delta_{ab}
~~~~~~~~~ \mathrm{for} ~~ A^{\mu}_{a} ~~ \subset P^{+} \nonumber ~, \\
&&\mathbb{V}\big(\{0|\pi R\}\big)_{ab}=\delta_{ab},~~\mathbb{V}\big(\{-1|0\}\big)_{ab}=(-1)\delta_{ab}
~~~ \mathrm{for} ~~ A^{5}_{a} ~~ \subset N^{+}~.
\end{eqnarray}

Now we present Lie algebra and its conventions in a specific canonical basis satisfying
\begin{eqnarray}\label{eq:commutation}
&&[\mathbf{C},\beta]=\beta E_{\beta},
~~[E_{\beta},E_{-\beta}]=\beta \cdot \mathbf{C} \nonumber \\
&&[E_{\beta},E_{\gamma}]
=N_{\beta,\gamma}E_{\beta+\gamma}~~\mathrm{if~\beta+\gamma\neq0}
\end{eqnarray}
where $\mathbf{C}$ is a vector of Cartan generators (or sometimes $C_{i}$ is denoted as a component of a Cartan generator), and $E_{\beta}$ is a root generator corresponding raising or lowering operator, and  $N_{\beta,\gamma}$ are structure constants. Note that the structure constants in this basis are not totally symmetric.
The roots of given original bulk gauge group $G$ denoted by $\alpha$ from now on can be chosen as generators of the $SU(2)$, whereas the generator of hypercharge  $Y$ can be a linear combination of Cartan generators denoted by $y\cdot \mathbf{C}$.
Then, we obtain the following relations,
\begin{eqnarray}
J_{0}=\frac{1}{|\alpha|^{2}}\,\alpha\cdot\mathbf{C},&&J_{+}
=\frac{\sqrt{2}}{|\alpha|}\,E_{\alpha},~~J_{-}=(J_{+})^{\dagger}\nonumber \\
&&Y=y\cdot \mathbf{C} ~,
\end{eqnarray}

\subsection{Weak mixing angle and Higgs mass in phenomenologically viable GHU models}
%\subsection{Weak mixing angle and effects of brane kinetic terms}
In order for GHU models to be phenomenologically viable, the following four requirements should be satisfied~\cite{Aranda:2010hn};
(1) all simple roots should be either isodoublets or isosinglets, (2) all isosinglets should have zero hypercharge, (3) all isodoublets should have the same non-zero hypercharge, (4) fermion representations with hypercharge 1/6 should be contained. Note that among these requirements, the fourth one reflects the spirit of the grand unified theories. Henceforth, we will only concentrate on the GHU models which can satisfy above four requirements.
When we consider those, it is sufficient to deal with simple roots only since we can always transform the root to one specific simple root among other simple roots by using suitable permutation and inversions of some axes.

Now, let us consider the general forms of the mass matrix of vector-boson divided by two types corresponding to
two generators $P^{+}$ and $N^{+}$,
\begin{eqnarray}
A_{\mu}=W_{\mu}^{+}E_{\alpha}+W_{\mu}^{-}E_{-\alpha}+W_{\mu}^{0}\hat{\alpha}\cdot
\mathbf{C}+B_{\mu}\,\hat{y}\cdot \mathbf{C}+\cdots ~~ \subset &&P^{+} \nonumber \\
A_{n}=\sum_{\beta >0} (\phi_{n,\beta}E_{\beta}+\phi_{n,\beta}^{*}E_{\beta})
+ \cdots ~~~~~~~~~~~~~~~~~~~~~~~~~~ \subset &&N^{+}. \label{An}
\end{eqnarray}
Here,  $A_{n}$ is a field corresponding to extra spatial components, and the $\phi$ fields correspond to 4D scalars.
Using Eq.(\ref{An}), the term $-\mathrm{tr} [A_{\mu},A_{n}]^{2}$ becomes
\begin{equation}
\sum_{\beta > 0~;~isodoublets} |\phi_{n,\beta}|^{2} \Big\{ \frac{1}{2}\alpha^{2}W^{+}_{\mu}W^{-\mu}
+\Big(W^{0}_{\mu}\hat{\alpha}\cdot\beta + B_{\mu}\,\hat{y}\cdot\beta\Big) \Big\}~. \label{An2}
\end{equation}
Supposing that the 4D scalar gets a vacuum expectation value from the effective potential and using
$\hat{\textit{y}}\cdot\beta=1/2$ and $s=\alpha \cdot \beta / |\alpha|^{2}=-1/2$,
the form of Eq.(\ref{An2}) is rewritten as
\begin{equation}
\sum_{\beta > 0~;~isodoublets} |\phi_{n,\beta}|^{2} \Big\{ \frac{1}{2}\alpha^{2}W^{+}_{\mu}W^{-\mu}
+\frac{1}{4}\alpha^{2}\Big(W^{0}_{\mu} - \frac{1}{|\alpha||\textit{y}|}B_{\mu}\Big)^{2} \Big\}~,
\end{equation}
and the weak mixing angle is given by
\begin{equation}
\tan{\theta_{W}}=\frac{1}{|\alpha||\textit{y}|}.
\end{equation}

Let us now consider the effect of the brane kinetic terms in the general case.
As learnt from the $SU(3)$ toy model described in the previous section, these brane kinetic terms we introduced can shift the coupling constants of $SU(2)$ and $U(1)$ in the same way. Re-scaling the gauge coupling constants, the term leading to the Higgs mass is written by
%\begin{equation}
%\sum_{\beta > 0~;~isodoublets} |\phi_{n,\beta}|^{2} \Big\{ \frac{1}{2}\frac{g^{2}}{Z_{1}}
%\alpha^{2}W^{+}_{\mu}W^{-\mu}+\frac{g^{2}}{4}\frac{1}{Z_{1}}\alpha^{2}W^{0\,2}
%-\frac{g^{2}}{2}\alpha^{2}\frac{1}{\sqrt{Z_{1}Z_{2}}}\frac{W\cdot B}{|\alpha||y|}+\frac{1}{4}\frac{g^{2}}{Z_{2}}\alpha^{2}
%\frac{B^{2}}{|\alpha|^{2}|y|^{2}} \Big\}~.
%\end{equation}
\begin{equation}
\sum_{\beta > 0~;~isodoublets} |\phi_{n,\beta}|^{2} \big(\frac{g^{2}}{Z_{1}}\big)
\Big[ \frac{1}{2}\alpha^{2}W^{+}_{\mu}W^{-\mu}
+\frac{1}{4}\alpha^{2}\Big(W^{0}_{\mu} - \sqrt{\frac{Z_{1}}{Z_{2}}}\frac{1}{|\alpha||\textit{y}|}B_{\mu}\Big)^{2} \Big]~,
\end{equation}
and the weak mixing relation is given by
\begin{equation}\label{eq:mixing angle}
\tan \theta_{W}^{\,\prime}=\frac{1}{|\alpha| |\textit{y}|}\,\sqrt{\frac{Z_{1}}{Z_{2}}}~.
\end{equation}
Similar to the case of 5D $SU(3)$ toy model, the weak mixing angle %$\Theta_W$%
is largely deviated from the experimental result ($\tan \theta_{W}^{~exp} \sim 1/\sqrt{3}$) even in the case of $c_{1}=c_{2}=0$
in any gauge group except for $E_{6}$ and $E_{8}$ as can be seen in Table.~\ref{tab:table1}.
In the GHU models, it is difficult to remedy this problem by incorporating the warped space time like RS1~\cite{Randall:1999ee} mainly due to the flat profiles corresponding zero modes as the SM gauge bosons before the EWSB.
Until now there have been two more well-known prescriptions to cure the probelm besides introducing the brane kinetic terms; one is to use non-simple Lie group like a product group, $G_{1}\times G_{2}$ or $G \times \sum_{i} U(1)^{i}_{A}$, where $U(1)^{i}_{A}$ means an anomalous U(1) that does have unclear origin in some cases, and the other is to let each original group live in different extra dimension that has different volume size, $M^{3,1}\times S^{1}\times (S^{1})^{\prime}$. Consequently those methods including brane kinetic terms can easily lead to the shift of each gauge coupling as much as we want.
%However above two methods are not suited in GUT spirit where we want to explain the well-known gauge unification from only one %simple group at very high energy.

\begin{figure}
\begin{center}
\includegraphics[width=0.6\textwidth]{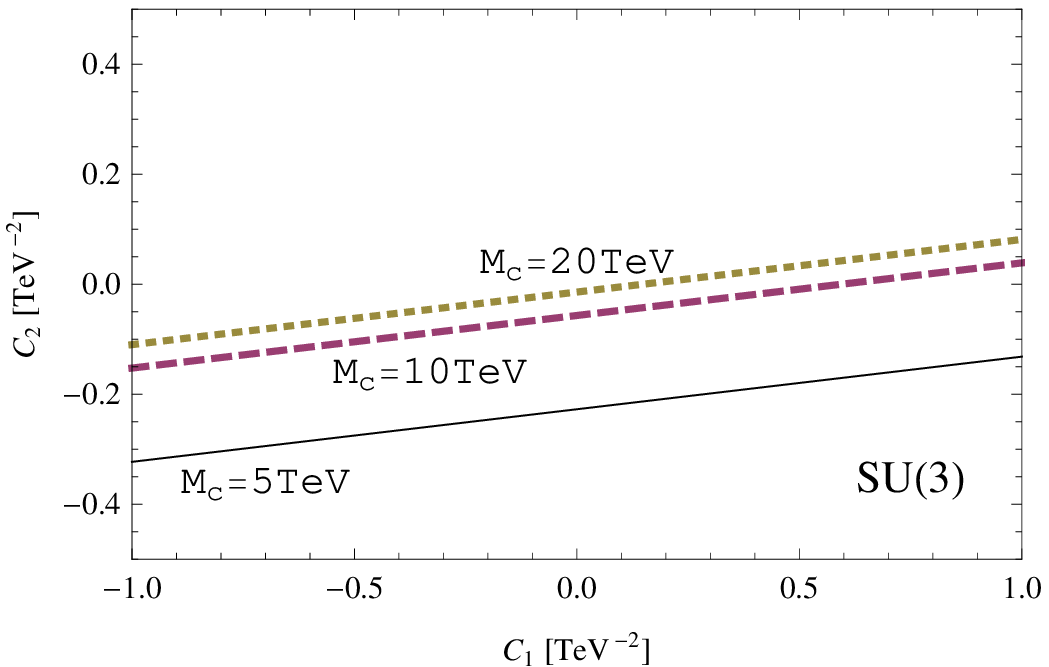}
\includegraphics[width=0.6\textwidth]{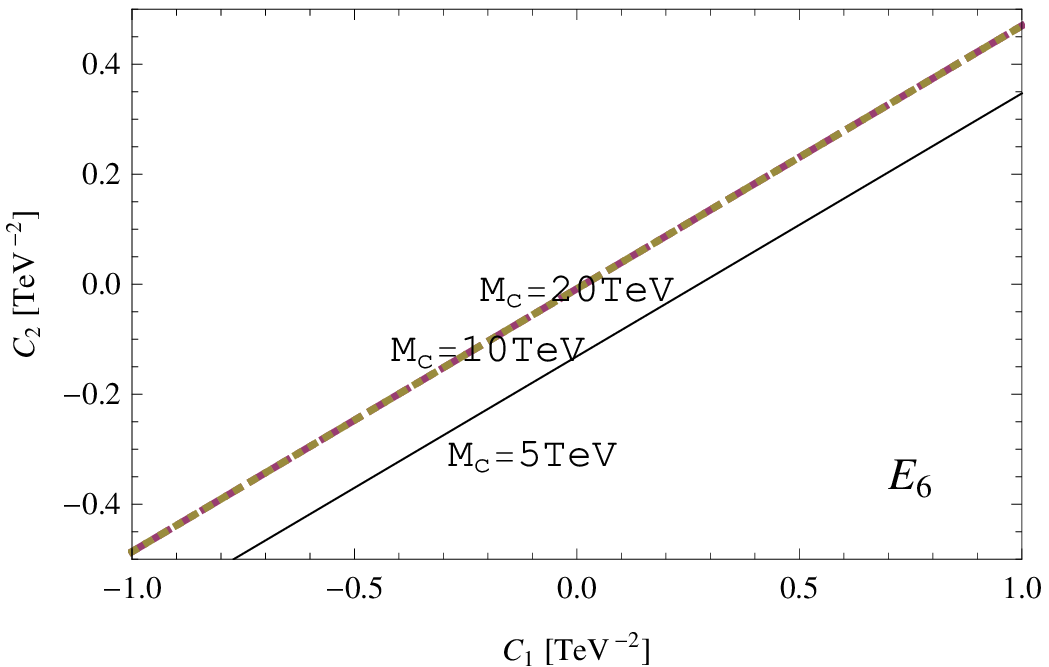}
\caption{ Correlations between  $c_1$ and $c_2$ for  $\tan \theta_{W}=0.535601$ \cite{Nakamura:2010zzi} and each group theoretic numerical factor in 6D $SU(3)$ and $E_{6}$ gauge Higgs unification models on $S^{2}/ \mathbb{Z}_{2}$. The straight, dashed and dotted lines correspond to the compactification scale $M_C$ to be 5, 10 and 20 TeV, respectively.}
\label{fig:ccp}
\end{center}
\end{figure}

Now, let us show that there exists a correlation between $c_{1}$ and $c_{2}$. Rewriting Eq.~(\ref{eq:mixing angle}) in terms of $c_{1}$ and $c_{2}$,
we can obtain the following relation
\begin{equation}
c_{1}=Z_{0}^{2}\Big(\frac{\tan^{2} \theta_{W}^{\prime}}{\tan^{2} \theta_{W}}-1\Big)+\frac{\tan^{2} \theta_{W}^{\prime}}{\tan^{2} \theta_{W}}\,c_{2}~.
\end{equation}
For a given group theoretic numerical factor, the values of $c_1$ and $c_2$ should be chosen to be consistent with the present experiment value of the weak mixing angle. The correlations between $c_{1}$ and $c_{2}$ in 6D $SU(3)$ and $E_{6}$ GHU models on $S^{2}/ \mathbb{Z}_{2}$ are shown in Fig.~\ref{fig:ccp}.
In the figure, the straight, dashed and dotted lines correspond to the compactification scale $M_C$ to be 5, 10 and 20 TeV, respectively.
We take $\tan \theta_{W}=0.535601$ as the experimental value~\cite{Nakamura:2010zzi}.
One can easy see from Fig.~\ref{fig:ccp} that the slope of the lines for $E_{6}$ is steeper than that for $SU(3)$ because the predicted value of $\tan\theta_W$ $(\sqrt{3/5})$ in $E_{6}$ is smaller than that $(\sqrt{3})$ in $SU(3)$, and the slopes of the lines are independent of the volume factor. The volume factors only change the value of $c_{2}$ depending on the scale $M_{C}$, as can be seen in both panels. Especially, the bottom panel shows that the scale dependence of the lines disappears for $M_C \gtrsim 10$ TeV in the case of $E_{6}$.

Finally, let us draw our attention to the Higgs mass.
We assume that the Higgs potential is given by
\begin{equation}
V(H)=-\mu^{2}|H|^{2}+\lambda|H|^{4}~.
\end{equation}
 As is well known, there are two methods for the Higgs scalar to get its mass. One is to use Wilson loops along extra dimensions. Especially, when the extra dimensional manifold is not simply connected, that is, $\Pi_{1}(K)\neq e$, the Wilson loop gets additional phases and the 4D fluctuations of the phase become the Higgs field. However, one-loop generated Higgs mass is in general too small, even smaller than W boson mass in this case. The other is to consider 6D theory with background fields. In this case, the quartic coupling of the Higgs scalar is naturally generated from higher dimensional non-Abelian gauge group, for instance, $\mathrm{tr}(F_{56}^{2})$. The negative quadratic term of the Higgs potential demanded to trigger the EWSB can be generated by the interaction between the original gauge field and background fields, $~\mathrm{tr}([A_{5},A_{B}])$. Note that the Higgs mass in latter case can be large enough because the terms in the Higgs potentail are generated at the tree level unlike the former case. So we necessarily consider the 6D models and even more higher dimensional models from now on.

After the Higgs scalar gets  vacuum expectation value, $<|H|>=v$, we can easily calculate the masses of Higgs and W boson, $M_{H}=\sqrt{2} \mu=\sqrt{2\lambda}v$ and $M_{W}=gv/2$, respectively. Thanks to the gauge kinetic terms, the ratio between the Higgs and W boson masses is given by
\begin{equation}
\frac{M_{H}}{M_{W}}=2\sqrt{Z_{1}}~, \label{WH}
\end{equation}
where we have  taked the quartic coupling $\lambda$ to be equal to $g^{2}/2$ \footnote{See the appendix for the details.}.
It is worthwhile to notice that the above relation is almost independent of how the $\mu$ term generates and the exact magnitude of the VEV.
In addition, extension to higher dimensional space does not affect the relation itself. It is only sufficient to trace gauge group breaking pattern from one simple group at the high energy to the SM gauge groups at the low energy, and estimate a simple volume factor of fundamental domain of the orbifolds.
Since $Z_{1}$ (or $c_{1}$) is written in term of $Z_{2}$ (or $c_{2}$) by
\begin{equation}
Z_{1}=\frac{\tan^{2} \theta_{exp}}{\tan^{2}\theta_{W}}\, Z_{2}~,
\end{equation}
the ratio given in Eq.(\ref{WH}) is rewritten by
\begin{equation} \label{eq:slope}
\frac{M_{H}}{M_{W}}=2\frac{\tan \theta_{exp}}{\tan \theta_{W}}\sqrt{ \Big( 1+\frac{c_{2}}{Z_{0}^{2}} \Big)}~.
\end{equation}
From the fact that all gauge couplings come from one gauge coupling in higher dimension, one can see why
the ratio only depends on one variable $c_{2}$.
Thus, one can easily achieve the experimental value of weak mixing angle by adjusting those two parameters $(c_{1}, c_{2})$ appropriately.
%%%%%%%%%%%%%%%%%%%%%%%%%%%%%%%%%%%%%%%%%%%%%%%%%%%%%%%%%%%%%%%%%%%%%%%%%%%%

\section{Numerical Results and Concluding Remarks}
We make numerical analysis in the GHU models with phenomenologically viable simple Lie groups satisfying four low energy constraints given in the previous section whose simple root $\alpha$ and member y of Cartan subalgebra correspond to the SM gauge groups $SU(2)$ and $U(1)$, respectively~\cite{Grzadkowski:2006tp}.
The numerical results for the weak mixing angle as well as the  Higgs mass are presented in Table \ref{tab:table1}. In the numerical calculation, we take $M_{W}=80.399$ GeV and $\sin^{2} \theta_{exp}=0.22292$ as inputs. \footnote{ See \cite{Nakamura:2010zzi} for the details about these experimental numbers.}
 As can be seen from Table I, the Higgs masses are predicted to be below the current lower bound the 114.4 GeV \cite{Barate:2003sz} in the case of $c_{2}=0$ for  all groups.

\begin{table}[h!t]
\caption{Simple Lie groups satisfying four low energy constraints~\cite{Grzadkowski:2006tp} and their predictions of the Higgs mass in the case of $c_{2}=0$, where $\alpha$ and $y$ correspond to $SU(2)$ weak and $U(1)$ hypercharge groups of the Standard model, respectively.} \label{tab:table1}
\begin{center} % put inside center environment
\begin{tabular}{|c||c|c|c|c|c|}
  \hline
  \hline
~~Group~~&~~ $\alpha$ ~~&~~~~ y ~~~~& $\tan \theta_{W} / \,\sqrt{\frac{Z_{1}}{Z_{2}}} $ ~~&~~~ Higgs mass [GeV], $c_{2}=0$ ~~~~\\
  \hline
  \hline
  $SU(3l)$   & $\alpha^{1}$     & $\tilde{\mu}_{2}/2$   & $\sqrt{3l/(3l-2)}$ & 49.7235 $\times \sqrt{(3l-2)/l}$    \\
  \hline
  $SO(2n+1)$ & $\alpha^{1}$     & $\tilde{\mu}_{2}/6$   & $\sqrt{3}$ & 49.7235     \\
  \hline
  $G_{2}$    & $\alpha^{1}$     & $\tilde{\mu}_{2}/6$   & $\sqrt{3}$  & 49.7235   \\
  \hline
  $F_{4}$    & $\alpha^{1}$     & $\tilde{\mu}_{2}/6$   & $\sqrt{3}$  & 49.7235   \\
  \hline
  $E_{6}$    & $\alpha^{1,5}$   & $\tilde{\mu}_{2,3}/2$ & $\sqrt{3/5}$ & 111.185   \\
  \hline
  $E_{7}$    & $\alpha^{1,7}$   & $\tilde{\mu}_{2,3}/6$ & $\sqrt{3}$,$\sqrt{3/2}$   & 49.7235, 70.3196  \\
  \hline
  $E_{8}$    & $\alpha^{1,8}$   & $\tilde{\mu}_{2,3}/6$ &$\sqrt{9/7}$,$\sqrt{3/5}$ & 75.9539, 111.185  \\
  \hline
\end{tabular}
\end{center}
\end{table}

\begin{figure}[h!t]
\begin{center}
\includegraphics[width=0.60\textwidth]{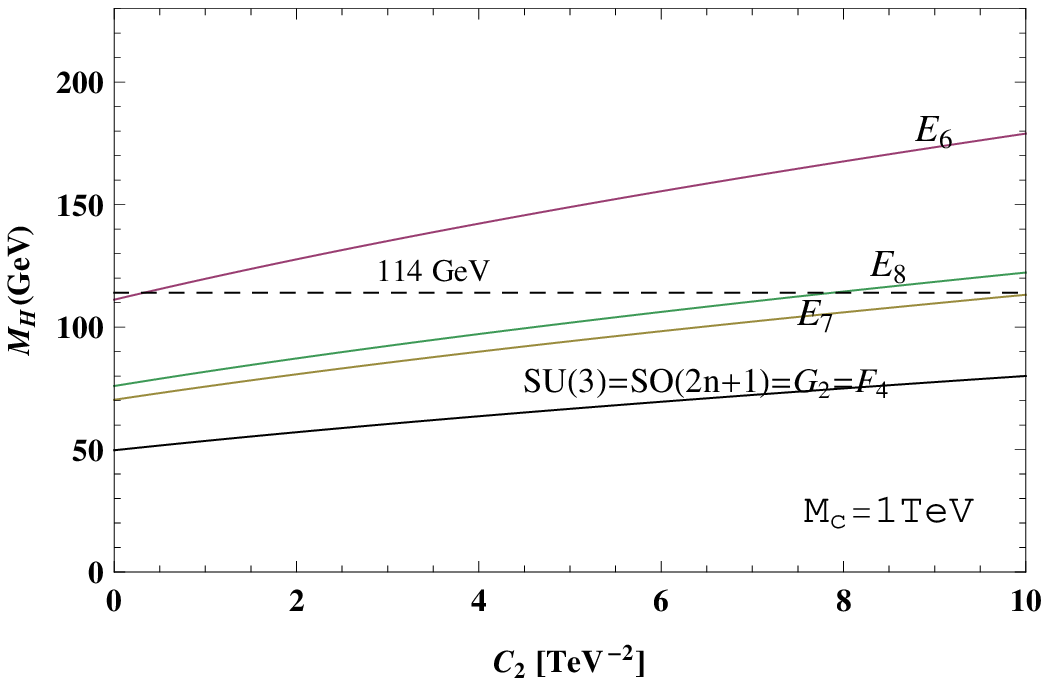}
\includegraphics[width=0.60\textwidth]{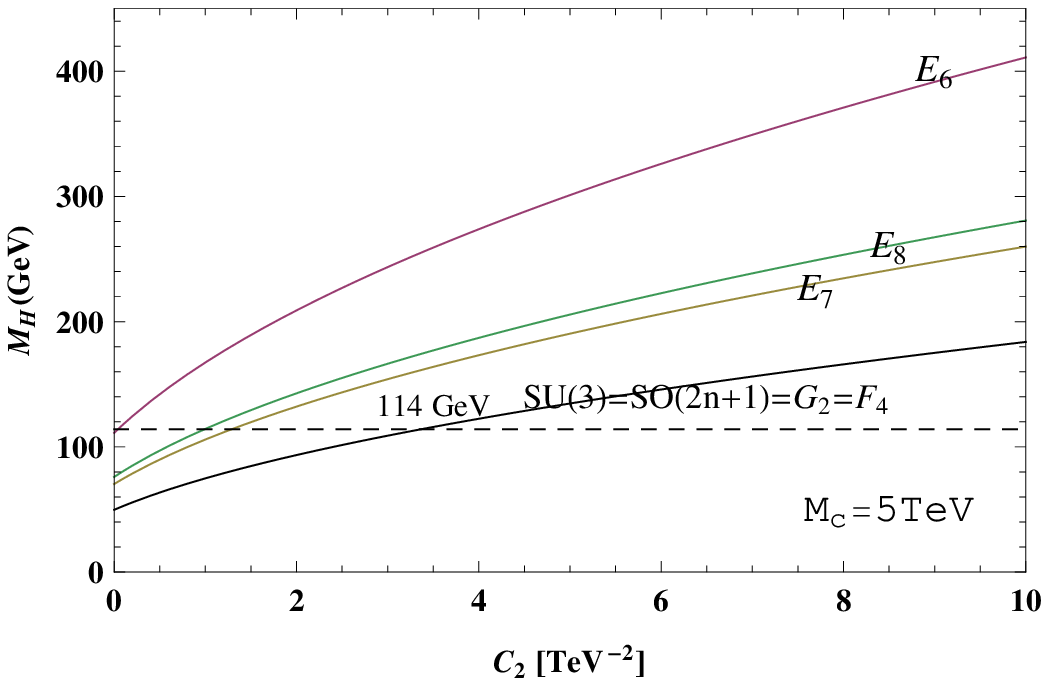}
\includegraphics[width=0.60\textwidth]{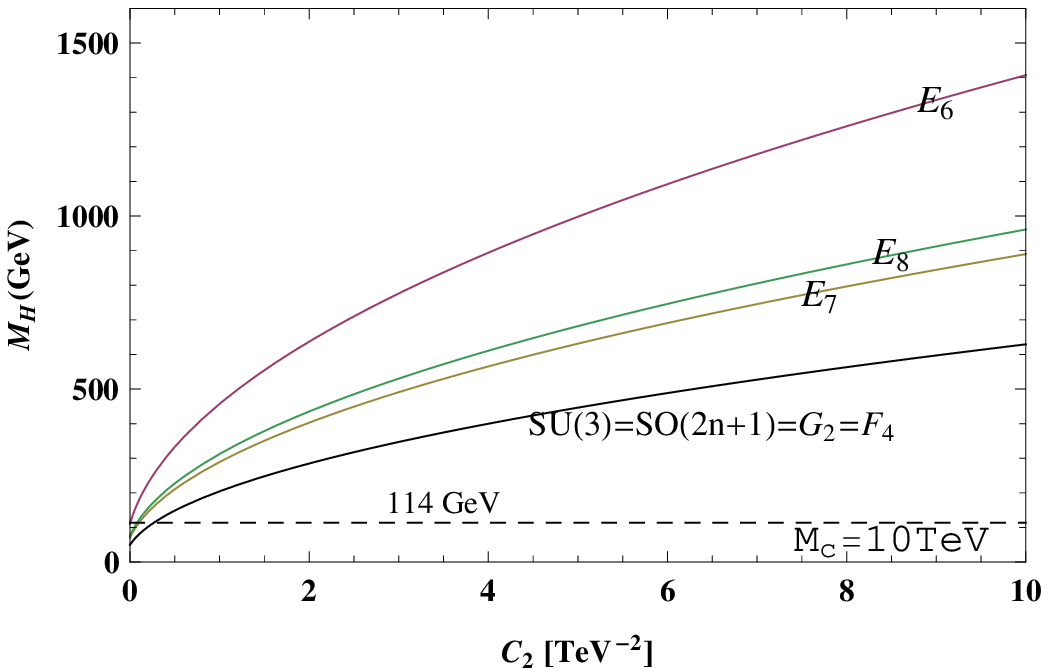}
\caption{Plots of the Higgs masses as a function of $c_{2}$ for several gauge groups   in the 6 D GHU model on $S^{2}/ \mathbb{Z}_{2}$. The compactification scale $M_C$ is taken to be 1 TeV (upper panel), 5 TeV (middle panel) and
10 TeV (lower panel).
}
\label{fig:figure1}
\end{center}
\end{figure}

\begin{figure}[h!t]
\begin{center}
\includegraphics[width=0.60\textwidth]{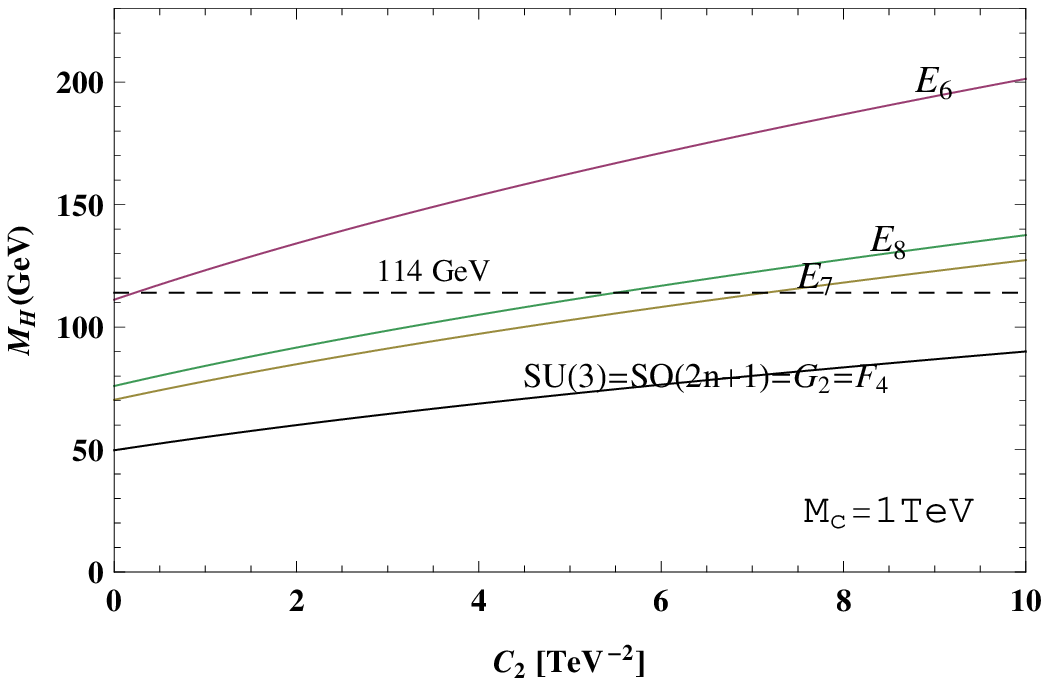}
\includegraphics[width=0.60\textwidth]{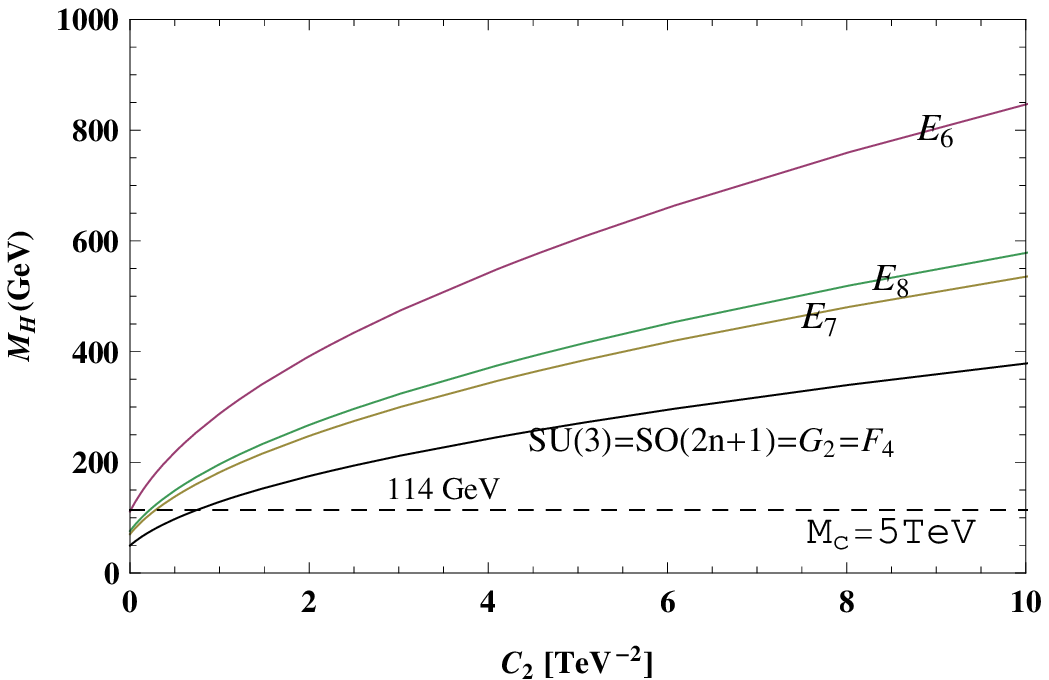}
\includegraphics[width=0.60\textwidth]{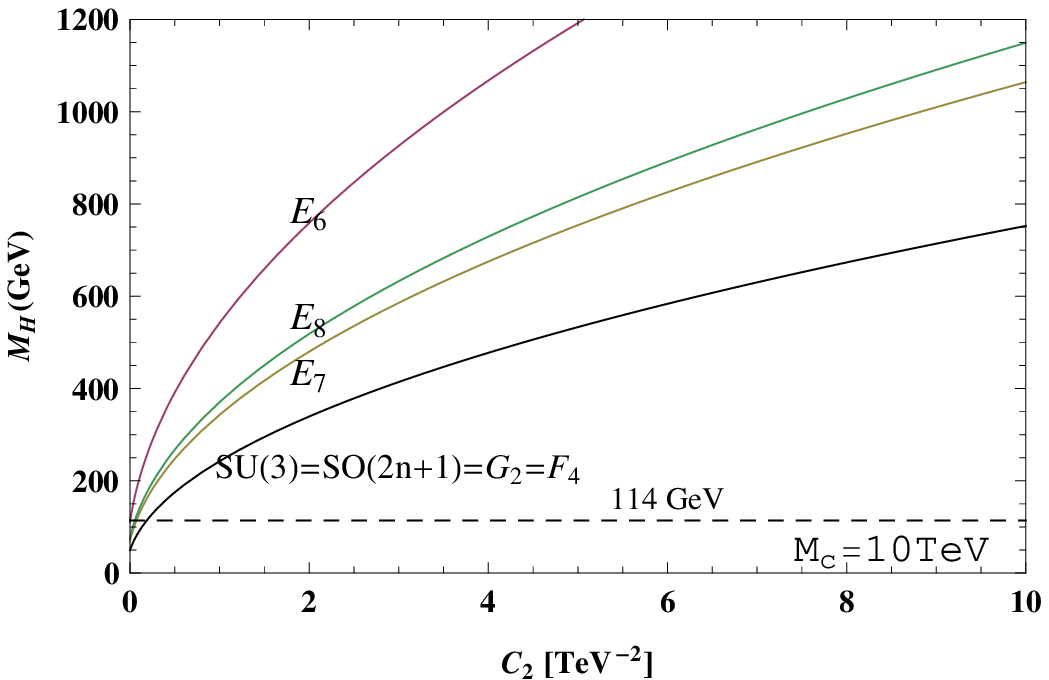}
\caption{Plots of the Higgs masses as a function of $c_{2}$ for several gauge groups   in the 6 D GHU model on $S^{2}/ \mathbb{Z}_{3}$. The compactification scale $M_C$ is taken to be 1 TeV (upper panel), 5 TeV (middle panel) and
10 TeV (lower panel)
}
\label{fig:figure2}
\end{center}
\end{figure}

For a given gauge group, we can estimate the Higgs mass as a function of $c_2$ by fixing the compactification scale $M_{C}$,  the structure of orbifold, and their dimension of extra space.
In Figs.\ref{fig:figure1} and \ref{fig:figure2}, we plot the Higgs mass as a function of $c_{2}$ for several allowed groups
and three different compactification scales $(M_{C})$
in 6 D GHU models with  $S^{2}/ \mathbb{Z}_{2}$ and $T^{2}/ \mathbb{Z}_{3}$ orbifold structure, respectively.
In the upper panel of Fig.\ref{fig:figure1}, we see that most values of the Higgs mass predicted for $M_C=1$ TeV in $E_{6}$ lie above the current experimental lower bound on the Higgs mass, 114 GeV,  whereas $c_2 \gtrsim 10$ is required for obtaining the Higgs masses consistent with the current lower bound in other exceptional groups $E_{7,~8}$. We also found that other groups like $SU(3l)$, $SO(2n+1)$, $G_{2}$, and $F_{4}$ can not lead to the Higgs mass consistent with the current lower bound unless
$c_{2}$ is taken to be very large.
As can be seen from the middle and lower panels of Fig.\ref{fig:figure1}, the situation can be substantially recast as $M_C$ gets increased. For $M_C=10$ TeV, we see that most predictions of the Higgs mass in all groups under consideration are consistent with the current lower bound.
Similarly, Fig.\,\ref{fig:figure2} shows similar behaviors as in the previous case of $S_{2}/ \mathbb{Z}_{2}$. In this case we choose $N=3$ as a specific example for the orbifold $T^{2}/ \mathbb{Z}_{N}$ because of the fact that the mod-out number $N$ is related to the number of the Higgs doublet, for instance,  there exist only two doublets for N=2,  only one doublet for N=3, and one doublet or no doublet for N=4,6 in a 6D torus case \cite{Scrucca:2003ut}. It is necessary to notice that only difference between Fig.1 and 2 is their different slopes of curves along with the scale $M_{C}$ and the structure of orbifold.
The reason why the slope of curve depends on the scale $M_C$ is that it becomes larger and larger as $M_{C}$ get increased, as can be seen from Eqs.~(\ref{eq:slope}) and ~(\ref{eq:slope2})~. The slope also strongly depends on the volume factor $Z_{0}^2$ of higher dimensional space and $\tan \theta_{W}$ calculated by purely group theoretical analysis after breaking  the original gauge group,
\begin{equation}\label{eq:slope2}
\frac{\partial M_{H}(c_{2})}{\partial c_{2}}=\frac{M_{W}}{Z_{0}^2}\frac{\tan \theta_{exp}}{\tan \theta_{W}}
\Big( \sqrt{1+ \frac{c_{2}}{Z_{0}^{2}}}~ \Big)^{-1/2}~.
\end{equation}
Note that the slopes of the curves do not depend on some details in high energy physics.

\begin{figure}[h!t]
\begin{center}
\includegraphics[width=0.6\textwidth]{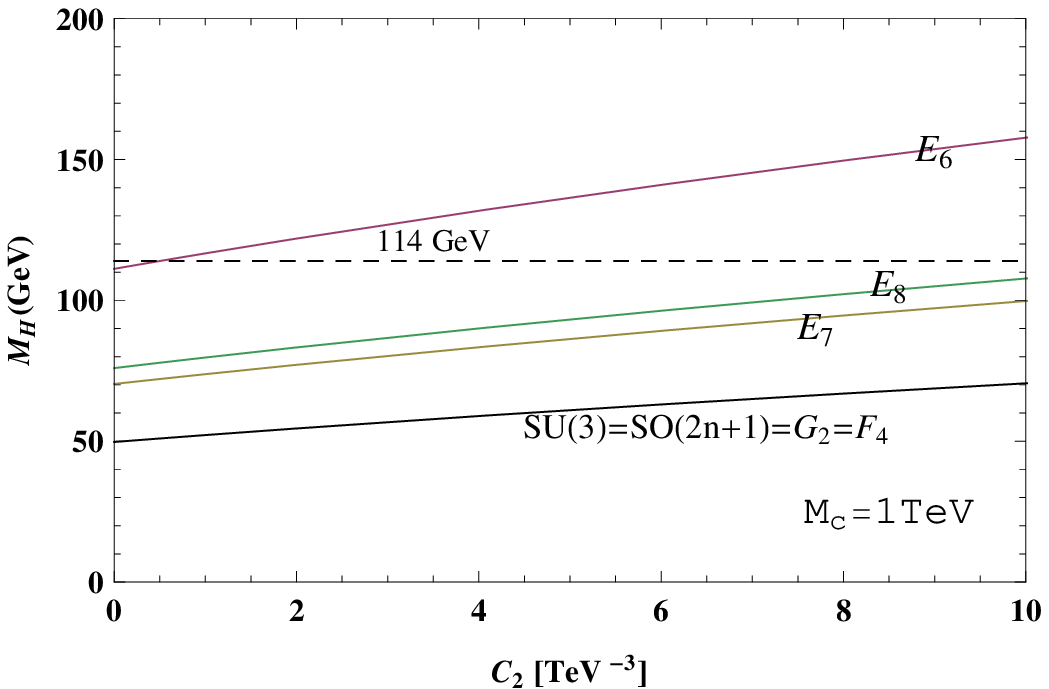}
\includegraphics[width=0.6\textwidth]{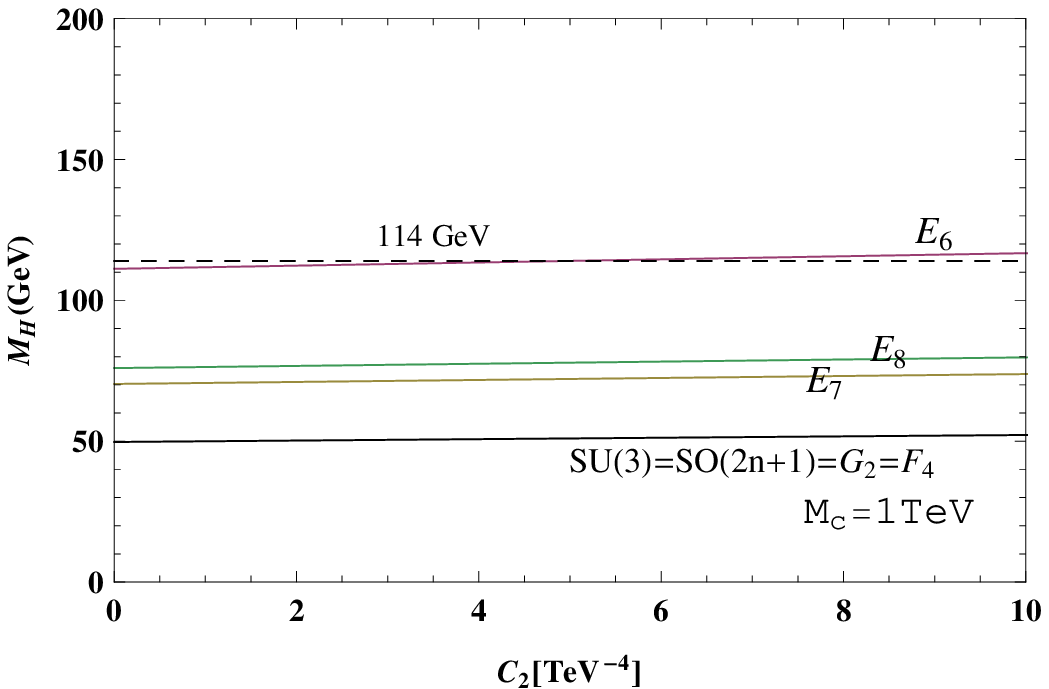}
\caption{The Higgs masses as a function of $c_{2}$ for various groups and $M_C=1$ TeV in the 7D GHU model on $S^{3}/Z_{2}$(upper panel) and 8D GHU model on $T^{4}/Z_{2}$(bottom panel).}
\label{fig:figure3}
\end{center}
\end{figure}

In Table \ref{tab:2}, we list several volume factors and numerical values of the slopes of the curves in the 6 D $E_{6}$ or $SU(3)$ GHU models on the $S^{2}/ \mathbb{Z}_{2}$ and $T^{2}/ \mathbb{Z}_{3}$, respectively,  in the 7 D $E_{6}$ GHU model on $S^{3}/ \mathbb{Z}_{N}$, and the 8 D $E_{6}$ GHU model on $T^{4}/ \mathbb{Z}_{N}$.
We also present how the values of the slope can be shifted with different choice of $c_{2}$ and the orbifold for given compactification scales $M_{C}=0.5, 1, 5$ TeV.
As examples, we take $c_{2}$ to be 0 and 10.
From Table \ref{tab:2}, one can see that the slope does not change largely in the cases of low $M_{C}$.
This is associated with why most predictions of the Higgs mass in the cases of low $M_C$ lie below the lower bound
on the Higgs mass 114 GeV.
In fact, it is not easy to make the Higgs mass to lie above the lower bound in many cases without taking a large value of $c_{2}$.
\begin{table}
\caption{Several volume factors and numerical values of the slopes of the curves in the 6 D $E_{6}$ or $SU(3)$ GHU models on the $S^{2}/ \mathbb{Z}_{2}$ and $T^{2}/ \mathbb{Z}_{3}$, and in the 7 D $E_{6}$ GHU model on $S^{3}/ \mathbb{Z}_{N}$ and the 8 D $E_{6}$ GHU model on $T^{4}/ \mathbb{Z}_{N}$. We take $c_{2}$ to be 0 and 10 to see how the results can be changed with the different choice of $c_{2}$ for given compactification scale $M_{C}=$0.5, 1, 5 TeV. In addition, $c_{2}$ corresponding to the Higgs mass to be 114.4 GeV is denoted by $c_{2}^{*}$.}\label{tab:2}
\begin{center} % put inside center environment
\begin{tabular}{|c|c|c|c|c|c|c|}
  \hline
  \hline
~Dimension~&~Space ~&~ Volume ~&~~~ $M_{C}$ ~~ &~ slope at $c_{2}=0$ &~slope at $c_{2}=10$&~~~~~ Remark ~~~~~\\
%  \hline
%  \hline
%  5D   & $S^{1}/ \mathbb{Z}_{N}$ & $2\pi R / N$  &  1 TeV &  &  &   \\
  \hline
  \hline
  6D  & $S^{2}/ \mathbb{Z}_{N}$ & $4 \pi R^{2}/ N $           &  0.5 TeV  & 2.212  & 2.195 & ~$N=2$,~$E_{6}$,~~$c_{2}^{*}=1.474$~~ \\
  \cline{4-7}
      &                        &                               &  1 TeV  & 3.957  & 3.534 &  ~$N=2$, ~$SU(3)$,~~$c_{2}^{*}=26.98$~~  \\
  \cline{4-7}
       &                         &                             &  1 TeV  & 8.848  & 7.903 &  ~$N=2$, ~$E_{6}$,~~$c_{2}^{*}=0.3685$~~ \\
  \cline{4-7}
       &                       &                             &  1 TeV & 17.70 & 12.47 &  ~$N=4$, ~$E_{6}$,~~$c_{2}^{*}=0.1843$~~  \\
  \cline{4-7}
       &                      &                             &  5 TeV & 221.2 &  17.52 &   ~$N=2$, ~$E_{6}$,~~$c_{2}^{*}=0.0147$~~  \\
  \cline{2-7}
       & $T^{2}/ \mathbb{Z}_{N}$ & $\big(2\pi R/ N \big)^{2}$  & 0.5 TeV   & 3.168 & 3.118 & ~$N=3$, ~$E_{6}$,~~$c_{2}^{*}=1.029$~~   \\
  \cline{4-7}
       &                         &                             &  1 TeV  & 5.668  & 4.598 &  ~$N=3$, ~$SU(3)$,~~$c_{2}^{*}=18.83$~~ \\
  \cline{4-7}
       &                         &                             &  1 TeV  & 12.67  & 10.28 &  ~$N=3$, ~$E_{6}$,~~$c_{2}^{*}=0.2573$~~ \\
  \cline{4-7}
       &                         &                             &  1 TeV  & 50.69  & 16.61 &  ~$N=6$, ~$E_{6}$,~~$c_{2}^{*}=0.064$~~ \\
  \cline{4-7}
       &                        &                             & 5 TeV  & 316.8 & 17.55 & ~$N=3$, ~$E_{6}$,~~$c_{2}^{*}=0.010$~~   \\
%       &                         &                             &         &  &  &  \\
%       &                         &                             &         &  &  &   \\
  \hline
  \hline
  7D   & $S^{3}/ \mathbb{Z}_{N}$ & $2\pi^{2} R^{3}/ N $  & 0.5 TeV & 0.7041 &  0.7035      & $N=2$, $E_{6}$,~~$c_{2}^{*}=4.632$~~  \\
  \cline{4-7}
       &                         &                       & 1 TeV & 5.633 &  5.364        & $N=2$, $E_{6}$ \\
  \cline{4-7}
       &                        &                       & 5 TeV & 704.1  &  17.57        & $N=2$, $E_{6}$ \\
  \hline
  \hline
  8D   & $T^{4}/ \mathbb{Z}_{N}$ & $\big(2\pi R/ N \big)^{4}$ & 0.5 TeV  & 0.0357 & 0.0357 & $N=2$, $E_{6}$,~~$c_{2}^{*}=91.43$~~  \\
  \cline{4-7}
       &                         &                            & 1 TeV  & 0.5707 & 0.5704 & $N=2$, $E_{6}$ \\
  \cline{4-7}
       &                     &                            & 5 TeV & 356.7 & 17.56 &   $N=2$, $E_{6}$ \\
  \hline
\end{tabular}
\end{center}
\end{table}
It is interesting that the slope of the curve becomes very small as the number of dimension increases.
Fig.~\ref{fig:figure3} shows how the predictions of the Higgs mass for the cases of the 7 D $S^{3}/ \mathbb{Z}_{2}$ and the 8 D $T^{4}/Z_{2}$ can change compared to those in the case of 6 D.
We see that only $E_{6}$ in the case of 7 D (the upper one) leads for  the Higgs mass to lie above the current lower bound on the Higgs masses, whereas other groups demand very large values of $c_2$ so as for the Higgs mass to be consistent with the current lower bound for $M_C= 1$ TeV. However, in the 8 D case, the Higgs masses for all groups weakly depend on the value of $c_2$ and lie below the lower bound.
Therefore, any 8 D GHU models with $M_c=1$ TeV cannot simultaneously accommodate the weak mixing angle and the Higgs mass coming from experiments without taking very large value of $c_{2}$ which looks rather unnatural. Consequently, we can conclude that rather large value of $M_{C}$ is necessarily demanded to accommodate the current  experimental data in higher-dimensional GHU models.
%\textbf{thanks to the observation the drastic change of the slopes for various scales $M_{C}$ above 1 TeV (see the fifth column of Table %\ref{tab:2}).}
%\textbf{But they can easily escape from the current bound since the slopes of the curves for various scales $M_{C}$ above 1 TeV in more higher dimensional models can change drastically.}

In summary, we have shown that the measured value of the weak mixing angle can be achieved by introducing the brane kinetic terms in GHU models.
%We have obtained a little bit modified relation between the Higgs and $W$ boson masses.
Interesting points that we found from our numerical results are  as follows;
\begin{itemize}
\item{A modified relation between the Higgs and $W$ boson masses is derived.}
\item{The current lower bound on the Higgs mass tends to favor exceptional groups $E_{6,~7,~8}$ than other $SU(3l)$, $SO(2n+1)$, $G_{2}$, and $F_{4}$ groups irrespective of the compactification scale $M_C$. The largest Higgs mass above the lower bound
    is achieved in $E_{6}$ among all groups we consider.}
\item{The predictions of the Higgs masses in the models for $SU(3l)$, $SO(2n+1)$, $G_{2}$, and $F_{4}$ groups can be consistent with the curent lower bound only when $M_C$ is higher than 1 TeV and $c_2$ is taken to be unnaturally large value.}
\item{For higher dimensional models such as 7D $S^{3}/ \mathbb{Z}_{2}$ and 8D $T^{4}/Z_{2}$,
      it is possible to get  the Higgs mass consistent with the current lower bound without taking unnaturally large value of $c_2$
      as long as  $M_C$ is taken to be higher than 1 TeV.}
\end{itemize}
Interestingly enough, the higher rank $E_{6}$ GHU model can be a remnant of one fundamental theory like the string theory whose gauge structure is $E_{8}\times E_{8}$ at very high energy scale, since $E_{6}$ GHU model with brane kinetic terms can be viable at even lower compactification scales below 1 TeV. Finally we anticipate that a great signal about the Higgs boson(or sector) detected soon at the Large Hadron Collider(LHC) would reveal the secret of electroweak symmetry breaking and the origin of the mass generation beyond the SM, and eventually judge which model has been chosen by Nature.

\appendix
\section{Generation of the quartic coupling of the Higgs in the 6D GHU}
We show how the Higgs gets a quartic coupling in the 6D GHU model. For our purpose, it is very useful to use a complex coordinate $z=(x^{5}+i x^{6})/\sqrt{2}$ and associated gauge field $A_{z}=(A_{5}-i A_{6})/\sqrt{2}$. The gauge field $A_{Z}$ can be written in terms of its zero modes and corresponding generators $E_{\beta~,~\gamma}$ by
\begin{equation}
A_{Z}=\frac{1}{2}h_{u}E_{\beta,\uparrow}+\frac{1}{2}h_{d}E_{\gamma,\downarrow}
+\frac{1}{2}h_{u}^{\prime}E_{-\beta,\uparrow}+\frac{1}{2}h_{d}^{\prime}E_{-\gamma,\downarrow}~,
\end{equation}
where $\uparrow$ and $\downarrow$ denote eigenvalues of the operator $s_{z}$ of the SM $SU(2)$, $\pm1/2$, respectively and thus $(h_{u},h_{d})$ and $(h_{u}^{\prime},h_{d}^{\prime})$ consist of each doublet member as follows;
\begin{equation}
H_{2}\equiv\frac{1}{\sqrt{2}}\left(
                          \begin{array}{c}
                            A_{z}^{(i)} -i A_{z}^{(i+1)}  \\
                            A_{z}^{(j)} -i A_{z}^{(j+1)}  \\
                          \end{array}
                        \right)
                        =\frac{1}{\sqrt{2}}\left(
                          \begin{array}{c}
                           h_{u}  \\
                           h_{d}  \\
                          \end{array}
                        \right)~,
\end{equation}
\begin{equation}
H_{1}\equiv\frac{1}{\sqrt{2}}\left(
                          \begin{array}{c}
                            A_{z}^{(i)} +i A_{z}^{(i+1)}  \\
                            A_{z}^{(j)} +i A_{z}^{(j+1)}  \\
                          \end{array}
                        \right)
                        =\frac{1}{\sqrt{2}}\left(
                          \begin{array}{c}
                           h_{u}^{\prime}  \\
                           h_{d}^{\prime}  \\
                          \end{array}
                        \right)~.
\end{equation}
Taking SU(3) as an example, we obtain
\begin{equation}
A_{Z}=\frac{1}{2}\left(
                   \begin{array}{ccc}
                     0 & 0 & A_{z}^{(4)} -i A_{z}^{(5)}  \\
                     0 & 0 & A_{z}^{(6)} -i A_{z}^{(7)} \\
                     A_{z}^{(4)} +i A_{z}^{(5)}  & A_{z}^{(6)} +i A_{z}^{(7)} & 0 \\
                   \end{array}
                 \right)~.
\end{equation}
 Let us suppose that there exists only one Higgs doublet at low energy and so we can set $H_{1}=0$. In order to obtain a quartic coupling,
 we consider $\mathrm{tr}[A_{Z},A_{Z}^{\dagger}]^2$.
The commutator, $[A_{Z},A_{Z}^{\dagger}]$, is calculated as follows;
\begin{eqnarray}
[A_{Z},A_{Z}^{\dagger}] &=& \frac{1}{4}\Big\{
                                    |h_{u}|^{2}[E_{+\beta}, E_{-\beta}]
                                    + h_{u}h_{d}^{\dagger}[E_{+\beta}, E_{-\gamma}]
                                    + h_{d}h_{u}^{\dagger}[E_{+\gamma}, E_{-\beta}]
                                    + |h_{d}|^{2}[E_{+\gamma}, E_{-\gamma}]
                                    \Big\}~, \nonumber \\
                            &=& \frac{1}{4}\Big\{
                                    |h_{u}|^{2}(\beta \cdot \mathbf{C})
                                    + h_{u}h_{d}^{\dagger}N_{+\beta,-\gamma}E_{+\alpha}
                                    + h_{d}h_{u}^{\dagger}N_{+\gamma,-\beta}E_{-\alpha}
                                    +  |h_{d}|^{2}(\gamma \cdot \mathbf{C})
                                    \Big\}~,
\end{eqnarray}
where we have omitted $\uparrow$ and $\downarrow$ subscripts and used the relations in Eq.~(\ref{eq:commutation}).
Here, note that $E_{\pm \beta(\gamma)}$ belongs to $N^{+}$, and thus the commutator between these two generators belongs to $P^{+}$
as can be seen from Eq.~(\ref{eq:NP})~\cite{Grzadkowski:2006tp}.
 With the relations among structure constants,
 $N_{\alpha,\beta}=-N_{\beta,\alpha}=-N_{-\alpha,-\beta}$ and orthonormalities of generators under the trace,
\begin{equation}
\mathrm{tr}C_{i}C_{j}=\delta_{i,j},~~~\mathrm{tr}E_{\alpha}E_{\beta}=\delta_{\alpha+\beta,0},~~\mathrm{tr}E_{\alpha}C_{i}=0~,
\end{equation}
Finally,
\begin{equation}
\mathrm{tr}[A_{Z},A_{Z}^{\dagger}]^{2}=\frac{1}{16}\Big\{
                                    |h_{u}|^{4}(\beta)^{2}
                                    + 2|h_{u}|^{2} |h_{d}|^{2}(\beta \cdot \gamma)
                                    + 2|h_{d}|^{2} |h_{u}|^{2} N_{+\beta,-\gamma}^{2}
                                    +  |h_{d}|^{4}(\gamma)^{2}
                                    \Big\}~. \label{comm}
\end{equation}
%Because we know which root in the root space is corresponding to the SM $SU(2)$ generator from Table.~\ref{tab:table1}, we can expect the $\beta$ for each group, so the $\gamma$ is also easily calculated from the fact that they are members of doublet, $\gamma=\beta - \alpha$.
 Using the relations $(\beta)^{2}=(\gamma)^{2}$, and $N_{+\beta, -\gamma}^{2}= -(\beta \cdot \gamma) + (\beta)^{2}$ \footnote{See Ref.~\cite{Lie-algebra} for the notation and properties of all root vectors} , we can further simplify Eq.(\ref{comm}),
\begin{eqnarray}
\mathrm{tr}[A_{Z},A_{Z}^{\dagger}]^{2} &=&\frac{1}{16}\Big\{|h_{u}|^{4}(\beta)^{2}  + 2|h_{u}|^{2} |h_{d}|^{2} (\beta)^{2}
+ |h_{d}|^{4}(\gamma)^{2} \Big\} \\ \nonumber
&=&\frac{1}{16}(\beta)^{2}\Big(|h_{u}|^{2} + |h_{d}|^{2} \Big)^{2} \nonumber \\
&=&\frac{1}{4}(\beta)^{2}(H_{2})^{4}~.
\end{eqnarray}
This result shows that the quartic coupling constant of the Higgs scalar is $1/4 \times (\beta)^{2}$.
From the magnitude of the $\beta$ root vector, $(\beta)^{2}=2$, the constant becomes $1/2$ as we introduced in main body of our paper.
%%%%%%%%%%%%%%%%%%%%%%%%%%%%%%%%%%%%%%%%%%%%%%%%%%%%%%%%%%%%%%%%%%%%%%%%%%%%%%%%%%%%%%%%%%%%%%%%%%%%%%%%%%%%
\newpage
\begin{center}
\acknowledgments
\end{center}
The work of S.K.Kang was supported in part by the National Research Foundation of Korea (NRF) grant funded by the Korea government of the Ministry of Education, Science and Technology (MEST) (No. 2011-0029758).
J. Park's work was supported by the Taiwan NSC under Grant No. 100-2811-M-007-030 and 099-2811-M-007-077.
J. Park also thanks J. S. Lee for the valuable comments and suggestions.
%%%%%%%%%%%%%%%%%% References %%%%%%%%%%%%%%%%%%%%%%%%%%%%%%%%%%%%%%%%%%%%%%%%%%%%%%%%%%%%%%%%%%%%%%%%%%%%%%%
%%%%%%%%%%%%%%%%%%%%%%%%%%%%%%%%%%%%%%%%%%%%%%%%%%%%%%%%%%%%%%%%%%%%%%%%%%%%%%%%%%%%%%%%%%%%%%%%%%%%%%%%%%%%%

\end{document}